\documentclass[a4paper]{ESASPCS13Style}
\usepackage{epsfig}

\begin{document}

\title{Photospheric temperature measurements in young main sequence stars}

\author{K. Biazzo\inst{1}, A. Frasca\inst{2}, G.\,W. Henry\inst{3}, S. Catalano\inst{2}\and E. Marilli\inst{2}} 
\institute{Universit\`a di Catania -- Dipartimento di Fisica e Astronomia, via S.\,Sofia 78, I--Catania, Italy 
\and INAF -- Osservatorio Astrofisico di Catania, via S.\,Sofia 78, I--Catania, Italy 
\and Tennessee State University -- Center of Excellence in Information Systems, 330 10th Ave. North, Nashville, 
TN 37203-3401}
\maketitle

\begin{abstract}
As part of our program to study stellar photospheric and chromospheric activity, we have examined several 
young solar type stars with activity levels intermediate between the Sun and the very active RS CVn 
binaries. We have analysed contemporaneous spectroscopic data obtained at Catania Observatory (Serra La 
Nave station, Mt. Etna) and photometric data acquired in the Str\"{o}mgren bands with an automatic 
photometric telescope (APT) at Fairborn Observatory (Arizona, USA). Surface inhomogeneities have been 
detected from the rotational modulation of stellar brightness as well as from the modulation of several 
photospheric line-depth ratios (LDRs). The presence of chromospheric plage-like regions has been inferred 
from the rotational modulation of the H$\alpha$ line equivalent width ($EW_{\rm H\alpha}$) evaluated with 
the spectral synthesis method. The most relevant results are the strong correlation between the brightness 
and temperature curves derived respectively from photometry and the LDRs as well as the striking 
anti-correlation between brightness and H$\alpha$ emission. This suggests a close spatial association of 
spots and plages, as frequently observed for the largest sunspot groups (e.g. \cite{Cata98}) and for some 
very active RS CVn systems (\cite{Cata02}). Moreover, a simple spot/plage model applied to the observed 
flux curves allows a rough reconstruction of photospheric and chromospheric  features of young main 
sequence stars.


\end{abstract}

\section{Introduction}
The simultaneous study of photospheric and chromospheric active regions on the Sun allows us to trace the 
emersion of magnetic flux tubes. Recently, a tight spacial association between spots and plages has been 
observed in the young solar type star HD 206860 by means of Str\"{o}mgren photometry and Ca{\small II} 
H$\&$K and H$\alpha$ chromospheric emissions (\cite{Fra00}). The spot/plage association has been also 
monitored in some very active RS~CVn binaries (\cite{Cata00}).

In this work we show that the spatial spot/plage association is also observed in other G-K main sequence 
stars younger than the Sun. We detected evidence of photospheric inhomogeneities from light curves and 
temperature measurements obtained by means of the line-depth ratios (LDRs) method (\cite{Cata02}). The 
chromospheric inhomogeneities have been detected from the variation of the H$\alpha$ line equivalent width. 

The cases of $\epsilon$ Eri (HD 22049, K2V, $B-V$=0.88), HD~166 (K0V, $B-V$=0.75), $\chi$1 Ori (HD 39587, 
G0V, $B-V$=0.59) and $\kappa$1 Cet (HD 20630, G5V, $B-V$=0.68) are considered.

\section{Observations and reduction}
\subsection{Photometry}
The photometric observations have been carried out in the standard Str\"{o}mgren {\it ubvy} system with the 
T4 0.75~m Automatic Photoelectric Telescope at Fairborn Observatory in southern Arizona (USA), equipped 
with an EMI 9124QB photomultiplier detector. A complete discussion of photometry with this telescope can 
be found in \cite*{Hen99}. We analyzed data acquired from November 2000 to January 2001, taken 
contemporaneously with the spectroscopic observations.

\subsection{Spectroscopy}
Spectroscopic observations have been obtained during the same time interval as the photometry with the 
REOSC \'echelle spectrograph fed by the 91-cm telescope at Catania Astrophysical Observatory - {\it M. G. 
Fracastoro} station (Serra La Nave, Mt. Etna). The spectral resolving power of about 14\,000 has been 
obtained in the cross-dispersed configuration with the 79-lines/mm \'echelle grating as a main dispersing 
element. The spectra were recorded on a CCD camera equipped with a thinned back-illuminated SITe CCD of 
1024$\times$1024 pixels (size 24$\times$24 $\mu$m). The detector allows us to record five orders in each 
frame, spanning approximately 5850 to 6700~\AA. In this spectral region there are the H$\alpha$ line and 
several line pairs, whose depth ratios are suitable for effective temperature determination (\cite{Cata02}). 
The average signal-to-noise ratio ({\it S/N}) at continuum in the spectral region of interest was 200-500 
for the very bright standard stars and 100-200 for the target stars.

The spectra extraction was performed by using the {\sc echelle} task of IRAF\footnote{IRAF is distributed 
by the National Optical Astronomy Observatory, which is operated by the Association of the Universities for 
Research in Astronomy, inc. (AURA) under cooperative agreement with the National Science Foundation.} 
following the usual steps: background subtraction, division by a flat field spectrum (given by a halogen 
lamp), wavelength calibration using the emission lines of a thorium-argon lamp, aperture extraction and 
continuum fitting with a low order polynomial. Detailed information about the data reduction can be found 
in \cite*{Cata02}.

\section{Data analysis and results}
Temperature determinations of our target stars have been made measuring the depth ratio of several line 
pairs, selected in the spectral interval 6190-6280~\AA. Calibrations of individual LDR into temperature 
scale have been made through the observation of non variable stars of different spectral type. More 
information about this technique is given in \cite*{Cata02}.

Excess emission in the H$\alpha$ line that contributes to filling in the line cores of the target stars 
have been extracted by using the ``spectral synthesis" method (e.g. \cite{Fra94}). For each active star, we 
have used the spectrum of an appropriate inactive star of the same spectral type that was rotationally 
broadened and subtracted from each individual spectrum of the target star. The net H$\alpha$ equivalent 
width has been measured in such difference spectra by integrating the net emission profile.

\subsection{Rotational modulation}
Rotational modulation of photospheric line fluxes in solar-type main sequence stars is not always evident. 
In our small sample, however, we have a different situation, and we describe our results on individual 
stars below.

\subsubsection{$\epsilon$ Eri and HD~166}
\label{sec:eps_Eri}
$\epsilon$ Eri and HD~166 are two examples where the variations induced by the presence of spots and 
plages at photospheric and chromospheric levels, as shown by the effective temperature and the H$\alpha$ 
line flux, are evident.

$\epsilon$~Eri is a bright ($V$=3\fm73, K2V), nearby (3.3 pc) single late-type main sequence star that 
shows variability attributed to magnetic activity. We have acquired spectra of this star from November 2000 
to January 2001. Phases are computed from the initial epoch of 24\,51856.0 corresponding to the first 
observing date (November 7, 2000) and a rotational period of $P_{\rm rot}$ = 11$\fd$68 (\cite{Dona96}). 
The 
rotational period is somewhat variable because spots appear at different latitudes and rotation is 
latitude-dependent (\cite{Frey91}). We have used the average photometric period for the epoch of our 
observations. We find a fairly well-defined anti-correlation between the photospheric temperature curve 
($<T_{\rm eff}>$), deduced by averaging nine LDRs, and the net H$\alpha$ equivalent width curve, derived 
with the spectral synthesis method using 54 Psc (HD 3651, K0V, $B-V$=0.849) as template. The full amplitude 
of the $<T_{\rm eff}>$ variation is only 50 K, i.e. about 1\%, while the equivalent width excursion is 
about 33\%.

\begin{figure}  %[ht]
\begin{center}
\epsfig{file=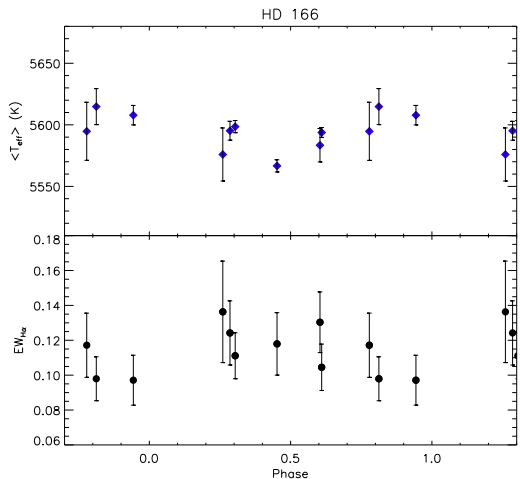, width=7.5cm}
\end{center}
\vspace{-3.cm}
\caption{Temperature variation curve (upper panel) obtained by using the line-depth ratios and H$\alpha$ 
emission curve (lower panel). 
\label{kbiazzof2}}
\end{figure}

HD~166 ($V$=6\fm13, K0V) is a young solar-type star belonging to the Local Association with an age between 
20 and 150 Myr. The data plotted in Fig.~\ref{kbiazzof2} refer to spectra acquired from October 2000 to 
November 2000. The $<T_{\rm eff}>$ and $EW_{\rm H\alpha}$ variations are reported as a function of the 
rotational phase, computed from the following ephemeris: HJD$_{\it \Phi =0}$ = 24\,49540.0 + 6$\fd$23 
$\times E$, taken from \cite*{Gai00}. The template used for the spectral synthesis method is $\tau$~Cet 
(HD~10700, G8V, $B-V$=0.727). Also in this case, the anti-correlation between photosphere and chromosphere 
is rather good. The full amplitude of $<T_{\rm eff}>$ variation is 48~K, i.e. about 1\%, while the 
$EW_{\rm H\alpha}$ excursion is about 34\%. 

\begin{figure}  %[ht]
\begin{center}
\epsfig{file=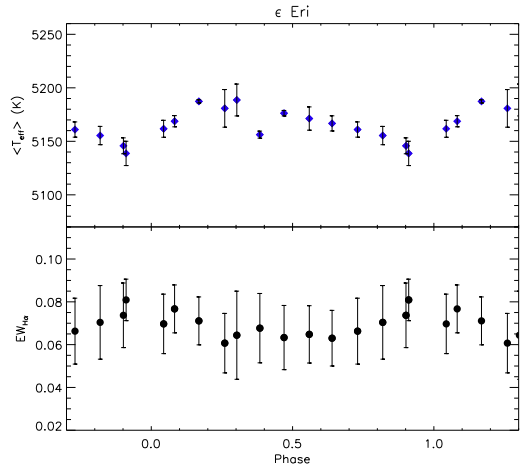, width=7.5cm}
\end{center}
\vspace{-1.5cm}
\caption{Temperature variation curve (upper panel) obtained by using the line-depth ratios and H$\alpha$ 
emission curve (lower panel).
\label{kbiazzof1}}
\end{figure}

\subsubsection{$\chi$1 Ori}
\label{sec:chi1_Ori}

\begin{figure}  %[ht]
\begin{center}
\epsfig{file=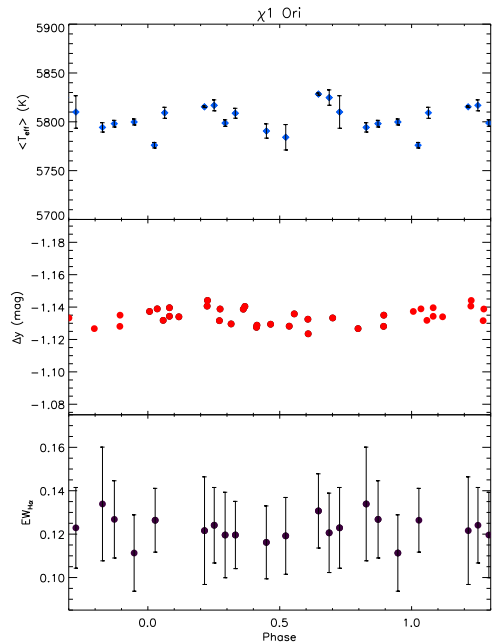, width=7cm}
\end{center}
\vspace{-.2cm}
\caption{Upper panel: Temperature variation curve obtained by using the LDR method. Middle panel: light curve 
given by Str\"{o}mgren differential photometry with HD 37147 (F0V, $B-V$=0.215) as comparison star. Bottom panel: 
H$\alpha$ equivalent width modulation in which 10 Tau (HD 22484, F9IV-V, $B-V$=0.574) has been used as template.
\label{kbiazzof3}}
\end{figure}

$\chi$1~Ori ($V$=4\fm41, G0V) is indeed a single-lined spectroscopic binary with a long period of 
$P_{\rm orb}$ = 5156$\fd$7 (\cite{Han02}). It is a relatively rapid rotator because it is a young star 
belonging to the Ursa Major Cluster with an age of 300 Myr. The simultaneous temperature, light and 
H$\alpha$ emission curves of this magnetically active star are reported in Fig. \ref{kbiazzof3}. The data 
have been folded in phase with the ephemeris HJD$_{\it \Phi =0}$ = 24\,51856.0 + 5$\fd$24 $\times E$, where 
the rotational period is taken from \cite*{Messi01}. The averaged effective temperature and the net 
H$\alpha$ equivalent width do not give clear evidence of rotational modulation, while the $\Delta y$ 
photometry displays a very low amplitude ($\sim$0.02 mag) modulation barely visible within the photometric 
noise. Since $\chi$1 Ori is a rather active star, as denoted by its large H$\alpha$ excess emission, we 
suppose that we have observed it at an epoch when the active regions were evenly distributed in longitude, 
which would give rise to the very low amplitude $y$ light and no H$\alpha$ modulation. The presence of 
significant H$\alpha$ filling and a mean magnitude lower than the historical maximum supports this picture.

Applying the analytical approach proposed by \cite*{Cata_bis02}, from the temperature variation amplitude 
we estimate a minimum spot coverage (with respect to the stellar surface) of about 0.0327 and 0.0275 for 
$\epsilon$~Eri and HD~166, respectively, where these two values have been obtained in correspondence of a 
ratio between spot temperature and photospheric temperature of 0.825. 

\subsubsection{k1 Cet and the Spot/Plage model}
\label{sec:spot_model}
$\kappa$1~Cet ($V$=4\fm83, G5V) is a single member of the Hyades moving group with an estimated age of $\simeq$750 
Myr. For this star we obtained simultaneous light and temperature curves, allowing us to make a spot model for a 
rough reconstruction of the photospheric inhomogeneities. We have also developed a plage model to be applied to 
the H$\alpha$ equivalent width curve. All curves show asymmetric shapes, so that all the solutions required at 
least two active longitudes.

\begin{figure}  %[ht]
\begin{center}
\epsfig{file=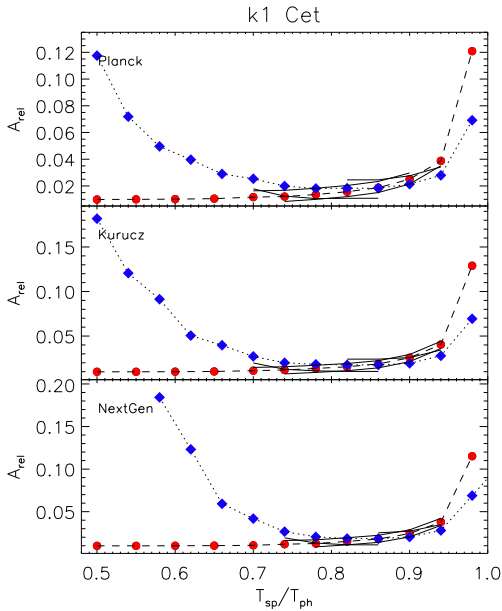, width=7.5cm}
\end{center}
\vspace{-2.8cm}
\caption{Grids of solutions for temperature curve (diamonds) and light curve (circles) obtained adopting the black 
body approximation and the two atmospheric models. The relative spot temperature and spot area we have found are 
$T_{\rm sp}/T_{\rm ph}$ = 0.855, 0.847, 0.863 and $A_{\rm rel} = \frac{A_{\rm spots}}{4 \pi R^2}$ = 0.0183, 0.0176, 
0.0184. The loci of reliable solutions in the $T_{\rm sp}/T_{\rm ph}$-$A_{\rm rel}$ plane are also marked in each 
panel.
\label{kbiazzof4}}
\end{figure}

The spot model that we used to reproduce the temperature and light curves is based on fixed geometric parameters 
of the cool spots (longitudes, latitudes) and solves for the spot relative area $A_{\rm rel}$, taking as a free 
parameter the ratio between the spot temperature and the photospheric temperature ($T_{\rm sp}/T_{\rm ph}$). In 
this way, we have obtained two grids of solutions, one for the $\Delta y$ curve and the other for the 
$\Delta<T_{\rm eff}>$ curve (Fig. \ref{kbiazzof4}), as we did in our previous work on RS CVn stars (\cite{Fra04}). 
Their intersection provides the best values of the spot temperature $T_{\rm sp}$ and the projected area of the 
spots relative to the stellar surface $A_{\rm rel}$. Moreover, for the evaluation of continuum flux, we have used 
the black body approximation and two atmospheric models developed by \cite*{Kuru93} and \cite*{Hau99}.

\begin{figure}  %[ht]
\begin{center}
\epsfig{file=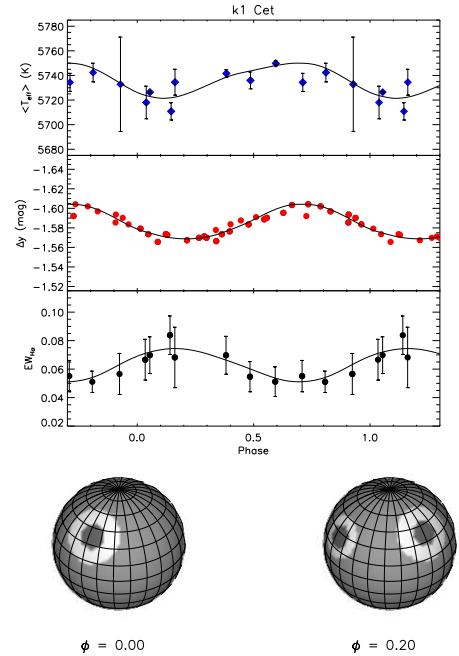, width=9cm}
\end{center}
\vspace{-.4cm}
\caption{Observed (dots) and synthetic (continuous lines) temperature, light and H$\alpha$ emission curves 
displayed as a function of the rotational phase. The template used for the $EW_{\rm H\alpha}$ measurements is 51 
Peg (HD 217014, G2.5IV, B-V=0.665). Schematic representation at two different phases of the photosphere and 
chromosphere of $\kappa$1 Cet, as reconstructed from the spot/plage model, are shown at the bottom of the figure.
\label{kbiazzof5}}
\end{figure}

For the H$\alpha$ curve we have considered a ``bright spot'' model with an emission flux ratio between plages and 
quiet chromosphere $F_{\rm pl}/F_{\rm ch}$=3, that is the typical value of the brightest solar plages.
In Fig. \ref{kbiazzof5} the $<T_{\rm eff}>$ variation, the $\Delta y$ photometry and the H$\alpha$ line flux of 
$\kappa$1 Cet are displayed as a function of the rotational phase (dots). The synthetic curves obtained by using 
the Kurucz model solutions are also shown. The ephemeris is HJD$_{\it \Phi =0}$ = 24\,51856.0 + 9$\fd$20 
$\times E$, where the rotational period is taken from \cite*{Gai00}. Both the temperature and light curves have a 
regular trend with the rotational phase and appear anti-correlated with the chromospheric flux curve. This implies 
a good spatial correlation between the stellar spots and the chromospheric plages. As a matter of fact, the 
photospheric and chromospheric maps have no appreciable longitude difference between the photospheric (spots) and 
chromospheric (plages) active regions. This result is similar to that obtained by \cite*{Fra00} for the young solar 
type star HD 206860.

\section{Discussion and conclusions}
\label{sec:conclusions}

The contemporaneous spectroscopic and photometric observations of some solar-type stars here reported have given 
remarkable results in the study of the connection between stellar photospheric and chromospheric active regions. 
The analysis of the rotational modulation of effective temperature and H$\alpha$ line equivalent width of our 
sample of active solar-like stars has really shown evidence of spacial correlation among the active regions, i.e. 
cool spots in photospheres and bright faculae in chromospheres in relatively low-activity stars.

Moreover, from the combined analysis of contemporaneous temperature and light curve variations, we have obtained 
unique solutions of the spot temperature and the spot coverage factor for the solar-type star $\kappa$1 Cet by 
using a method already tested in some RS CVn stars (\cite{Fra04}). Also, for $\epsilon$~Eri and HD~166, we have 
found a valuation of these two parameters. The spot temperatures we derived are closer to solar spot penumbrae 
rather than umbrae, probably due to the larger size of stellar penumbra, which has a greater weight in determining 
the spectral line depths. In addition, the relative spot coverage in these stars is much smaller than in the 
spotted RS CVn stars (\cite{Fra04}) and more similar to the largest solar spot groups.

\begin{acknowledgements}
We want to thank the SOC for the financial assistance.  GWH acknowledges support from NASA grant NCC5-511 and NST 
grant HRD-9706268.
\end{acknowledgements}


\begin{thebibliography}{}

\bibitem[\protect\astroncite{Catalano et al.}{2002a}]{Cata02}Catalano S., Biazzo K., Frasca A., Marilli E. 2002, 
A\&A, 394, 1009
\bibitem[\protect\astroncite{Catalano et al.}{2002b}]{Cata_bis02}Catalano S., Biazzo K., Frasca A. et al. 2002, 
AN, 323, 260
\bibitem[\protect\astroncite{Catalano et al.}{1998}]{Cata98}Catalano S., Lanza A.F., Brekke P. et al. 1998, in ASP 
Conf. Ser.: The 10th Cambridge Workshop on Cool Stars, Stellar Systems and the Sun, eds. Donahue R.A. \& Bookbinder 
J.A., vol. 154, p. 584
\bibitem[\protect\astroncite{Catalano et al.}{2000}]{Cata00}Catalano S., Rodon\`o M., Cutispoto G. et al. 2000, in 
Kluwer Academic Publishers: Variable Stars as Essential Astrophysical Tools, ed. Ibanoglu C., vol. 544, p. 687
\bibitem[\protect\astroncite{Donahue et al.}{1996}]{Dona96}Donahue R.A., Saar S.H., Baliunas S.L. 1996, ApJ, 466, 
384
\bibitem[\protect\astroncite{Frasca et al.}{2004}]{Fra04}Frasca A., Biazzo K., Catalano S. et al. 2004, A\&A, in 
press
\bibitem[\protect\astroncite{Frasca \& Catalano}{1994}]{Fra94}Frasca A. \& Catalano S. 1994, A\&A, 284, 883
\bibitem[\protect\astroncite{Frasca et al.}{2000}]{Fra00}Frasca A., Freire Ferrero R., Marilli E., Catalano S. 
2000, A\&A, 364, 179
\bibitem[\protect\astroncite{Frey et al.}{1991}]{Frey91}Frey G.J., Grim B., Hall D.S. et al., AJ, 102, 1813
\bibitem[\protect\astroncite{Gaidos et al.}{2000}]{Gai00}Gaidos E.J., Henry G.W., Henry S.M. 2000, AJ, 120, 1006
\bibitem[\protect\astroncite{Han \& Gatewood}{2002}]{Han02}Han I. \& Gatewood G. 2002, PASP, 114, 224
\bibitem[\protect\astroncite{Hauschildt et al.}{1999}]{Hau99}Hauschildt P.H., Allard F., Ferguson J. et al. 1999, 
ApJ, 525, 871
\bibitem[\protect\astroncite{Henry}{1999}]{Hen99}Henry G.W. 1999, PASP, 111, 845
\bibitem[\protect\astroncite{Kurucz}{1993}]{Kuru93}Kurucz R.L. 1993, ATLAS9 Stellar Atmosphere Programs and 2 
km s$^{-1}$ grid, (Kurucz CD-ROM No. 13)
\bibitem[\protect\astroncite{Messina et al.}{2001}]{Messi01}Messina S., Rodon\`o M., Guinan E.F. 2001, A\&A, 366, 
215

\end{thebibliography}
\end{document}